\documentclass[journal]{IEEEtran}

 \usepackage{cite,url}
 \usepackage{amsfonts,amssymb,verbatim}

\usepackage{amsmath,amssymb,graphicx,bm,color,amsbsy,amsthm,algorithm,algorithmic}
\usepackage{epstopdf}
\usepackage{nicefrac}

\newcommand{\bA}{{\mathbf{A}}}
\newcommand{\ba}{{\mathbf{a}}}
\newcommand{\bC}{{\mathbf{C}}}
\newcommand{\bH}{{\mathbf{H}}}
\newcommand{\bh}{{\mathbf{h}}}
\newcommand{\bI}{{\mathbf{I}}}
\newcommand{\bn}{{\mathbf{n}}}
\newcommand{\bw}{{\mathbf{w}}}
\newcommand{\by}{{\mathbf{y}}}

\begin{document}

\title{Distributed Base Station: A Concept System for Long-Range Broadband Wireless Access}

%

\author{Muhammed~Faruk~Gencel,~\IEEEmembership{Student Member,~IEEE,}
        Maryam~Eslami~Rasekh,~\IEEEmembership{Student Member,~IEEE,}
        and~Upamanyu Madhow,~\IEEEmembership{Fellow,~IEEE}
\thanks{The authors are with the Department of Electrical and Computer Engineering, University of California, Santa Barbara, CA, 93106 USA. E-mail:
\{gencel, rasekh, madhow\}@ece.ucsb.edu.}
}

\maketitle

\begin{abstract}
We propose a concept system termed distributed base station (DBS), which enables distributed transmit beamforming at large carrier wavelengths to achieve significant range extension and/or increased downlink data rate, providing a low-cost infrastructure for applications such as rural broadband. We consider a frequency division duplexed (FDD) system, using feedback from the receiver to achieve the required phase coherence. At a given range, $N$ cooperating transmitters can achieve $N^2$-fold increase in received power compared to that for a single transmitters, and feedback-based algorithms with near-ideal performance have been prototyped. In this paper, however, we identify and address key technical issues in translating such power gains into range extension via a DBS. First, to combat the drop in per-node SNR with extended range, we design a feedback-based adaptation strategy that is suitably robust to noise. Second, to utilize available system bandwidth, we extend narrowband adaptation algorithms to wideband channels through interpolation over OFDM subcarriers. Third, we observe that the feedback channel may become a bottleneck unless sophisticated distributed reception strategies are employed, but show that acceptable performance can still be obtained with standard uplink reception if channel time variations are slow enough. We quantify system performance compactly via outage capacity analyses.

\end{abstract}


\IEEEpeerreviewmaketitle

\section{Introduction}

\begin{figure} [h]
\includegraphics[width=1\columnwidth]{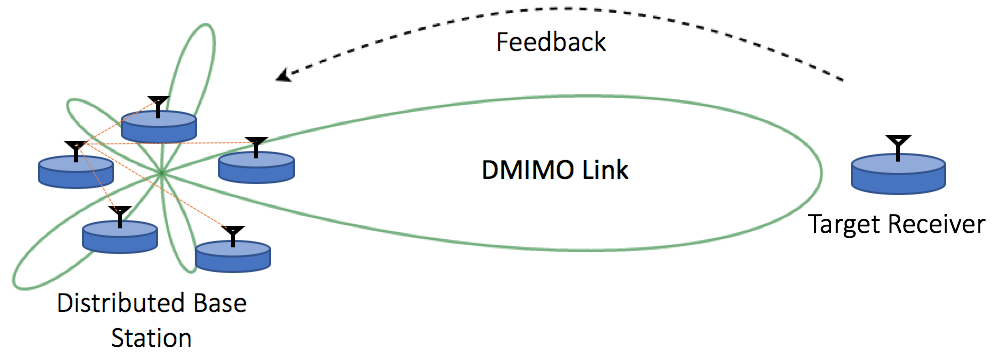}
\caption{The Distributed Base Station Concept System.}
\label{fig:scenario}
\end{figure}

Distributed transmit beamforming with $N$ cooperating transmitters can provide received power $N^2$-fold larger than that of a single transmitter. Over the past decade, there has been significant progress in demonstrating the feasibility of attaining the precise carrier frequency and phase synchronization required to realize these gains \cite{MudumbaiTWC07,mudumbai2005scalable,QuitinTWC13,quitin2016scalable}. In this paper, we build on these advances for design of a concept system that we term distributed base station (DBS), targeting significant improvements in communication link range and/or data rate.  As shown in Figure \ref{fig:scenario}, a DBS comprises $N$ opportunistically placed, low-cost, transmitter nodes, without wired connections between the nodes.  Our goal is to leverage the $N^2$-fold received power gain provided by distributed transmit beamforming to significantly
enhance downlink range and/or spectral efficiency.  While a DBS can be employed to enhance communication in existing WiFi and LTE bands, the approach is particularly interesting for white space frequencies (e.g., 50-800 MHz). These frequencies propagate well, and are therefore well matched to long-range applications such as rural broadband. However, good propagation also leads to poor spatial reuse when employing omnidirectional transmission.  Beamforming using multiple antennas can enhance spatial reuse, but centralized antenna
arrays are bulky at large wavelengths.  DBS allows the use of low-cost transmit nodes with moderate transmit power to emulate a powerful transmitter with a highly directional steerable antenna.

For $N$ coordinating transmitters, we can obtain an $N$-fold power pooling gain even without beamforming. In practice, this would require loose timing coordination between the transmitters (e.g., so the delay spread seen by the receiver across transmitters is smaller than the size of an OFDM cyclic prefix). While this addition is noncoherent, and hence leads to fading on narrowband channels, this effect is alleviated by frequency diversity in wideband systems.  Thus, we envision a system which starts with an $N$-fold power pooling gain, and then adapts to higher spectral efficiency as the DBS trains its beam towards the desired receiver, with the maximum spectral efficiency corresponding to the $N^2$-fold ideal beamforming gain.  As an example of the system enhancements possible with a $10$-node DBS (our running example), consider a rural broadband link served by a single transmitter node in white space frequencies (50-800 MHz). For a receiver at the cell edge, which could only sustain a very low rate control channel at $-5$ dB SNR with a single base station transmitter, a $10$-node DBS boosts the SNR at the receiver to 5 dB with power pooling, and to 15 dB with ideal beamforming. For typical SNR versus modulation and coding scheme (MCS) values for LTE adaptive modulation \cite{sesia2011lte}, we can sustain QPSK with rate $2/3$ coding with power pooling, and 64QAM with rate $2/3$ coding with beamforming. Thus, for a channel bandwidth of 20 MHz, a DBS makes it possible to provide broadband data rates of 20-80 Mbps at a range where a single transmitter system could barely establish a control link. It is worth noting, however, that in order to achieve these gains, the distributed beamforming scheme must operate effectively in regimes where the received SNR for any given transmit node is low.  Thus, we cannot use, for example, the well-known one-bit feedback based beamforming strategy originally proposed in \cite{mudumbai2005scalable}, which has been shown to be fundamentally limited in its performance for low per-node SNR \cite{gencel2015distributed}.

In terms of the range extension possible with an $N$-node DBS, a free space propagation model with inverse square decay of power with distance would predict an $N$-fold range increase, corresponding to the $N^2$-fold power gain due to beamforming.  The gains when
we use more realistic propagation models are smaller, but still significant.  The power of the DBS concept lies in our ability to harness MIMO at lower carrier frequencies. As an example link budget using the Hata propagation model in Section \ref{sec:downlink} indicates, by using a white space frequency of 200 MHz, we can attain spectral efficiencies larger than 3 bps/Hz at a range of 14 km, using a 10-node DBS with a relatively small transmit power per node of 20 dBm (the emitted power of a typical WiFi node). If we now use a larger carrier frequency of 800 MHz, the range drops to about 6 km.

\noindent
{\bf Contributions:} We propose and evaluate a wideband system design that addresses the key technical hurdles for realizing the DBS concept. 
We consider an FDD system in which the receiver sends feedback that enables the transmitters in the DBS to synchronize.  In order to attain protocol-level scalability, we constrain the feedback to be {\it aggregate} (i.e., not directed at any specific transmit node), so that the receiver is oblivious to the number and identity of the transmit nodes.  The key contributions are as follows:\\
1) We design a feedback-based synchronization strategy, which we term deterministic orthogonal sequence training (DOST). We explore its properties in
a narrowband setting, comparing it against a number of previously proposed strategies to show that it is better matched to the low per-node SNR regime of interest to us.  We also show that DOST is resilient to severe quantization on the feedback link, which as we discuss shortly, can become a bottleneck in the DBS system.\\
2) We show that DOST extends naturally to wideband frequency-selective channels by considering an OFDM system in which training sequences are
sent on a designated set of pilot subcarriers. The receiver feeds back a corresponding sequence of (quantized) received complex amplitudes, which the transmitters use to estimate their channels on the pilot subcarriers. These are then interpolated across subcarriers by each transmitter, and are used to beamform on the data subcarriers.  Assuming a feedback delay which is smaller than the time constant of channel variations, this approach scales to arbitrarily low per-node SNRs: measurement noise is averaged out effectively over a period scaling inversely with the per-node SNR. \\
3) We compactly characterize the downlink performance for a DBS via an outage capacity analysis for a well-accepted 3GPP channel model.\\
4) We note that feedback can become a bottleneck to varying degrees, with the rate of channel time variations that can be supported depending on the sophistication of the uplink reception strategy.  We use an outage capacity analysis to compare the channel coherence times that the DBS can support with and without distributed reception on the uplink.

{\bf Organization:} We put our work in context by briefly summarizing related work in Section \ref{sec:related}. In Section \ref{sec:system_model}, we describe the overall system design, and provide some example results.  In Section \ref{sec:approaches}, we 
describe the DOST scheme tor feedback-based training, and provide performance evaluations for a single subcarrier, including the effect of drastic feedback quantization.  
We justify our design choices by comparing DOST with alternative 
techniques such as one-bit feedback-based beamforming \cite{mudumbai2010distributed} and time-multiplexed training  \cite{thibault2013design}. In Section \ref{sec:wideband}, we provide results
showing the performance of DOST for a wideband system, showing that simple interpolation across the pilot subcarriers to estimate the channel for the data subcarriers works well.
Section \ref{sec:outage} provides an outage capacity analysis that compactly characterizes the coded performance of the wideband system, both for the downlink
and for the feedback on the uplink. Section \ref{sec:conclusion} contains our conclusions.

\section{Related work} \label{sec:related}
With the increasing availability of low-cost front-end elements, architectures that take advantage of the degrees of freedom offered by massive deployment of antenna elements have gained in popularity. Acquisition of adequate channel state information at the transmitter (CSIT) is crucial for FDD massive MIMO systems and has been of particular interest in literature with traditional codebooks for channel feedback \cite{love2003grassmannian, au2007performance} which requires the number of feedback bits to scale linearly with the number of BS antennas \cite{jindal2006mimo}. Efficient codebook design based on the channel statistics \cite{shen2017performance} and sparsity inspired approaches are proposed in \cite{shen2017high,lee2015antenna} to reduce feedback overhead. The fundamental differences between this body of work and our framework are as follows.
First, in order for the network protocols to scale with the number of distributed transmitters, and to allow opportunistic expansion of the DBS, we constrain the receiver to be oblivious to the number and identity of transmitters. Thus, instead of performing channel estimation and then producing quantized feedback, the feedback must be based on the receiver's {\it aggregate} measurements. This still allows us to consider standard training strategies, in which different transmitters send orthogonal training sequences, but constrains the form of feedback the receiver can send back.  This implies, for example, that the receiver cannot perform spatial channel estimation followed by codebook-based quantization, or exploit sparsity, unlike in existing feedback-based techniques in massive MIMO.  Second, the impact of operating in the low per-node SNR regime has not been considered in prior work on massive MIMO feedback.  As shown in this paper, this limitation on the feedback rate can fundamentally limit the channel coherence times that can be supported.

A simple example of distributed beamforming with aggregated feedback is the one bit feedback algorithm \cite{mudumbai2005scalable2,mudumbai2010distributed}, in which transmitters use small random phase perturbations to perform stochastic ascent on the received signal strength, based on the receiver feedback, broadcast to all transmitters, of one bit 
per iteration.  This approach is simple and has formed the basis for several prototypes \cite{QuitinTWC13,Quitin2013Receive}. The relatively slow convergence of the original algorithm
(e.g., $5N$ iterations to reach 75\% of ideal beamforming gain \cite{mudumbai2005scalable2}) can be improved to a limited extent by strategies such as
exploiting knowledge of previous perturbations \cite{song2010improving}, or using knowledge of channel statistics at the receiver \cite{thibault2010random}.
However, the method is fundamentally mismatched to low per-node SNRs \cite{gencel2015distributed}, roughly speaking, because noise masks the effect of small phase perturbations.  
This motivates our approach, in which the transmitters employ large phase perturbations during a training phase, rather than attempting to make small adjustments while beamforming.

The one-bit feedback algorithm has been extended to wideband regimes in a prior paper by the authors \cite{gencel2015scaling}, by adding an additional bit per subcarrier which enables enforcement of phase continuity across subcarriers.  This is fundamentally different from the training-based approach in this paper, which enables explicit channel estimation on pilot subcarriers for each transmitter, and hence is amenable to standard interpolation across subcarriers.

Once we commit to a training phase, one possible approach is for the transmitters to take turns transmitting in training slots, with the receiver sending (quantized) feedback corresponding
to each slot. Such time-multiplexed training has been successfully prototyped \cite{bidigare2012implementation}, and has been studied with quantized feedback in \cite{thibault2013design}.
The algorithm is energy efficient, since only one node is active per iteration, but unlike the DOST scheme proposed in this paper, it does not utilize integration over time to combat noise, 
so that its performance suffers in the low per-node SNR regime, as we show in our numerical results.  The time-multiplexed approach also does not scale well at the protocol level because of the dependence of the training frame structure on the number of transmitter nodes, and the coordination required between them to take turns.

Our work, and the related work discussed above, falls into the category of all-wireless distributed beamforming with explicit feedback, which allows flexible deployment of DBS supporting
FDD operation. 
It is worth mentioning recent work on all-wireless distributed beamforming based on channel reciprocity, that relies on tight pre-synchronization of the cooperating nodes to emulate
a centralized array with a common time base \cite{bidigare2015wideband,peiffer2016experimental,peiffer2016experimentalMilcom}.  
Finally, there is also significant recent work on distributed MIMO based on coordination of infrastructure nodes (WiFi access points or cellular base stations) via a fast wired backhaul \cite{irmer2011coordinated,balan2013airsync,rahul2012megamimo,hamed2016real}. Our emphasis is to scale up the number of nodes without fast wired backhaul and achieve massive MIMO gains \cite{larsson2014massive,choi2014downlink,rusek2013scaling}.

The present paper builds on our prior conference paper \cite{gencel2015noise}, which introduced the DOST scheme, and also discussed extensions to wideband systems.  However, it goes well beyond \cite{gencel2015noise} by presenting a DBS concept system built around DOST, including an OFDM system design roughly consistent with LTE, 
specific prescriptions for pilot design, and outage capacity analyses for downlink and uplink which compactly characterize system-level performance.

\section{System Model} \label{sec:system_model}

The nodes in the DBS depicted in Figure \ref{fig:scenario} cooperate to send a common message to a distant receiver over a noisy multipath channel.  We consider OFDM
with a set of subcarriers designated as pilots.  The receiver broadcasts explicit aggregate feedback on the complex signals received on the pilot subcarriers.
Each transmitter in the DBS uses this feedback to estimate its complex channel gains to the receiver on the pilot subcarriers, and interpolates these to estimate the
channel gains on the data subcarriers.  Each transmitter then adjusts its phase on each subcarrier to compensate for the channel phase, in order to align coherently
at the receiver.

The transmitters are assumed to be synchronized in time and frequency, and the channel is assumed to be quasi-static (i.e., it can be modeled as time-invariant with respect to the
time constant of the feedback-based channel estimation strategy). Timing synchronization is relatively straightforward with the OFDM system model because timing offsets smaller than the length of the cyclic prefix are tolerable.  As long as each transmitter maintains a fixed framing with respect to its own clock, drifts between the transmitters' clocks can be
compensated for by the feedback-based phase synchronization.  Frequency synchronization can be achieved by each node synchronizing to the receiver node (as done in prototypes
such as \cite{quitin2012distributed,QuitinTWC13}) or to a master node within the DBS, and the effect of small residual frequency offsets is similarly compensated for by the phase compensation algorithm.

We now describe the signal and channel models employed in the paper. We discuss and justify our design choices in more detail in later sections of the paper.  

\subsection{Signal Model}

We denote the channel from node $i$ to the receiver on subcarrier $k$ by the complex gain $H_i(f_k)= a_{ik} e^{j\psi_{ik}}$, and the receiver's phase offset relative to transmitter $i$ 
by $\gamma_{ik}$. Transmitter $i$ applies phase control via a beamforming weight of $e^{j\theta_{ik}}$ on its $k$th subcarrier. 
The received signal, after multiplying by the conjugate of the unit-amplitude pilot symbol,
is given by
\begin{equation} \label{eq:complex_sample}
 	R[f_k] = \sum_{i=1}^{N}{a_{ik} e^{j(\theta_{ik} +\gamma_{ik}+\psi_{ik})}}+ w[k]
\end{equation}
The corresponding normalized received signal strength (RSS) is given by
\begin{equation} \label{eq:normalized_rss}
 	 r[f_k] = \frac{\left| R[f_k] \right|}{N}
\end{equation}
and is used as a performance metric to compare different beamforming algorithms.

The aim of distributed transmit beamforming is to maximize RSS. This is achieved when each transmitter chooses a beamforming phase that reverses its total offset relative to the receiver allowing all signals to combine coherently upon reception. The optimal solution is therefore $\theta_{ik}=-(\gamma_{ik} + \psi_{ik})$ up to a common constant shift across all nodes. The received RSS is then equal to
\begin{equation} \label{eq:ideal_bf}
R_\text{max}[f_k] = \sum_{i=1}^{N}{a_{ik}}
\end{equation}
and the normalized RSS $r[f_k]$ approaches the maximum achievable value.

\subsection{Channel Model}

The multipath channel between a typical transmitter node and the receiver is modeled as
$$
h(\tau) = \sum_{p=1}^{N_p} \alpha_p \delta(\tau - \tau_p)
$$
where $N_p$ denotes the number of paths, $\tau_p$ the delay and $\alpha_p$ the complex gain of path $p$. 
For concreteness, we consider Rayleigh fading on each path, setting $\alpha_p = A_p v_p$ where $v_p$ are i.i.d. complex Gaussian with distribution $CN(0,1)$ and $A_p$ is the normalized square root of the power delay profile (PDP) such that $\sum_{p} A_p^2 = 1$.

The frequency response for such a channel is:
$$H(f_k) = \sum_{p} A_p v_p e^{-j 2 \pi f_k \tau_p}$$
For each frequency $f_k$, $H(f_k)$ is zero-mean complex Gaussian with variance $\sum_{i} A_p^2=1$.  Thus, the channel responses of each transmitter at different frequencies are identically distributed but correlated random variables with distribution  $H(f_k) \sim CN(0,1)$.

In our numerical results, we use the 3GPP Extended Pedesterian A (EPA) channel model parameters shown in Table-\ref{table:EPA} to generate channels for each node in the DBS.
Different nodes therefore have the same power-delay profile, but different channel realizations corresponding to i.i.d. draws of the $\{ v_p\}$. We note that this channel model
is not intended to provide a physical model of multipath components, but rather, may be viewed as a non-uniform tapped delay line representation of a bandlimited channel.  We have also considered dithered versions of the delays for different nodes, and verified by simulations that the channel statistics in frequency domain do not change. Thus, the channel model should
be viewed as a non-uniform tapped delay line representation, rather than a model for physical multipath components.

\begin{table}[h!]
\begin{center}
\begin{tabular}{|c| |c| |c| |c|}
 \hline
 \multicolumn{4}{|c|}{Parameters} \\
 \hline
 Path Number& Delay (ns)& Relative Power (dB) & Fading\\
 \hline
1&0    &0.0 &Rayleigh\\
\hline
2&30  &-1.0&Rayleigh\\
\hline
3&70 &-2.0& Rayleigh\\
\hline
4&90 &-3.0& Rayleigh\\
\hline
5&110&-8.0& Rayleigh\\
\hline
6&190&-17.2& Rayleigh\\
\hline
7&410&-20.8& Rayleigh\\ 
 \hline
\end{tabular}
\end{center}
\caption{EPA channel model \cite{3gpp.36.104}}
\label{table:EPA}
\end{table}

\section{Per-subcarrier feedback-based distributed beamforming  \label{sec:approaches}}

In this section, we describe and justify our choice of training strategy 
through numerical comparison against alternative approaches. 
We focus on a narrowband system which serves as a model for
a single pilot subcarrier in the wideband OFDM framework. As mentioned, we are interested in techniques that scale well, in terms of both performance and protocol simplicity, as the number of transmitter nodes increases
and as the received SNR per node approaches zero.

\subsection{Deterministic Orthogonal Sequence Training}

In this scheme, each node uses a predefined sequence of beamforming weights over $L$ training transmissions and the $L$ complex gains measured by the receiver are
quantized and broadcast to the transmitters. By using orthogonal or quasi-orthogonal weight sequences on different nodes, each node can extract its channel from the feedback independently from other nodes.

Consider the $L\times N$ training matrix $\bA$, the $i$'th column of which is the weight sequence used by node $i$. To design $N$ orthogonal sequences, each sequence must be at least of length $N$, meaning the training period, which is equivalent to convergence time for the iterative approaches, is equal to $N$ and  scales linearly with array size. Using orthogonal sequences is then equivalent to choosing a training matrix for which $\bA ^H \bA$ is diagonal. This orthogonality requires some coordination between transmitters to ensure one-to-one assignment of sequences to nodes. This requirement can be relaxed by using quasi-orthogonal pseudorandom sequences that nodes generate independently;
the normalized interference between sequences gets attenuated as their length grows. 
While it is possible to employ completely uncoordinated training by the latter choice, in practice, the coordination required for implementing truly orthogonal sequences (which provide the best 
possible performance for a given training duration and power) is minimal.  There are many possible choices of training sequences, but for concreteness, we consider the DFT matrix in our results:

\begin{equation*}
\bA = \begin{bmatrix}
1 & 1 & \cdots & 1 & 1\\ 
1 & e^{-j2\pi/L} &\cdots  &  & e^{-j2\pi(N-1)/L} \\ 
\vdots & \vdots & \ddots  &  & \vdots\\ 
1 &  &  &  & \\ 
1 & e^{-j2\pi(L-1)/L} & \cdots &  & e^{-j2\pi(N-1)(L-1)/L}
\end{bmatrix}
\end{equation*}
This is because DFT sequences are not only orthogonal, but they remain orthogonal when cyclically shifted by any amount. Thus, a transmitter can use any $L$-sized block of feedback to estimate its channel, without incurring interference from the sequences sent by the other transmitters. A receiver can therefore snoop on the pilot subcarriers at any time, and generate
feedback for the transmitter nodes in the DBS.  Similarly, any transmitter node can join or leave the DBS at any time, assuming basic OFDM frame alignment and frequency synchronization is maintained. This makes deployment and operation particularly simple.

The observations at the receiver, collected over times $l=1,...,L$, can be written as the $L \times 1$ vector
\begin{equation*}
\by = \bA \bh + \bw
\end{equation*}
where $\bh$ is the $N \times 1$ channel vector across different transmitters and $\bw \sim CN(0,N_0 \bI)$ is the receiver noise.

The least squares estimate for the channel vector is given by
\begin{equation}
\widehat{\bh}=\left( \bA^H \bA \right)^{-1} \bA^H \by
\label{eq:least_squares}
\end{equation}
assuming that $L \geq N$ and $\bA$ has rank $N$. 
Each node can thus obtain its channel estimate by taking the inner product of its corresponding row in the matrix $(\bA^H \bA)^{-1} \bA^H$ and the channel measurement feedback vector.
 
The Cramer-Rao lower bound on error covariance is $\bC_h = N_0 \left( \bA^H \bA \right)^{-1}$.
For each transmitter, the error covariance is bounded as
 $$\mathrm{Var}(\hat{\bh}_n) \geq (N_0(\bA^H \bA)^{-1})_{n,n} \geq  \frac{N_0}{(\bA^H \bA)_{n,n}}$$ 
 with the bound attained for orthogonal training (diagonal $\bA^H \bA$). In this case, each node can estimate its
 channel by separately correlating the observations with its own training sequence:
\begin{equation}
{\hat{\bh}}_n = \frac{1}{L} \ba_n^H \by = \bh_n + \frac{1}{L} \ba_n^H \bw
\label{eq:channel_est}
\end{equation}
where $\ba_n$ is the $n$th column of the training matrix.
The estimation error covariance $\mathrm{Var}(\hat{\bh}_n)=N_0/L$ can be made arbitrarily small by increasing the training
interval $L$. This also demonstrates the power-pooling advantage of the DOST algorithm; with $L$ scaling linearly with $N$, the estimation accuracy improves as array size grows and longer links with lower RSS can be supported.

\subsubsection*{Quantization}

In practice, the complex received signal amplitude measured at the receiver must be quantized to a limited number of bits and broadcast by the receiver. The variance of the received complex amplitude scales as $N$ (the transmitted signals add up incoherently during
the training period), hence a natural question is whether the quantization resolution also needs to be enhanced as $N$
increases.  Fortunately, the answer is no: as long as the receiver scales its quantizer step size $\Delta$ as $\sqrt{N}$ 
to accommodate the amplitudes it is seeing,
we can use a fixed number of quantization bins, and average out the quantization noise across the training period.

The channel estimate at transmitter $n$ with quantized feedback can be written as 
\begin{align*}
\hat{\bh}_n  & = \frac{1}{L}\ba_n^\bH (\by + \bn_q) \\
& = \frac{1}{L}\ba_n^\bH \by + \frac{1}{L}\ba_n^\bH \bn_q
\end{align*}
where $\bn_q$ is the quantization noise vector. Assuming quantization noise is distributed uniformly over the span associated to each level, the variance of any element $n_q[l]$ of the quantization noise vector scales as 
$$ \mathrm{Var}(n_q[l])=\frac{\Delta^2}{12} \sim N. $$
If the quantization noise values can be approximated as independent over time, we have
$$
\mathrm{Var}(\frac{1}{L} \ba_n^H \bn_q) \sim \frac{NL}{L^2} = \frac{N}{L}
$$
so that the effect of quantization noise on channel estimation can be made independent of $N$ by scaling $L$ linearly with $N$.
Thus, we can use a fixed feedback rate even as we increase the number of transmitters $N$, as long as the length of the training
period scales linearly with $N$.

\subsection{Alternative Strategies}

The goal here is not to be
comprehensive, but to show that the proposed DOST strategy is better matched to the low-SNR regime than methods that have been suggested in the literature, and natural
variants thereof. To this end, we consider per-node deterministic training, the one-bit feedback algorithm, and variants of the latter that employ 2 bits per iteration.

\subsubsection{Successive Deterministic Distributed Beamforming }
A special case of deterministic orthogonal training is time-multiplexed training where only one transmitter is active at a time and the phase offset measured at the receiver is fed back to individual nodes successively. This procedure, termed 
Successive Deterministic Distributed Beamforming (SDDB) in \cite{thibault2013design}, 
corresponds to setting the training matrix to identity, i.e., $\bA= \bI_{NxN}$. As our analysis and numerical results demonstrate, this method is poorly matched to the low per-node SNR regime of interest to us, since a transmitter is not able to use the
entire training period to average out noise.  Of course, if the per-node SNR is large enough SDDB may be preferable in terms of power conservation as only one transmitter is active at any time during a training period of similar length. 
This may be the case for shorter range applications, in which the goal of distributed beamforming is to reduce transmitted power rather than to obtain range
extension for a given transmitted power. 
On the other hand, SDDB is more resilient to quantization than DOST, since the dynamic range of the received signal is smaller when a single node is transmitting at a time.

\subsubsection{One bit feedback}
The one bit feedback algorithm (OBF) is a simple randomized iterative procedure that aims to synchronize the phase of  transmitters using a single bit of feedback for each measurement. This feedback is aggregated for all transmitters instead of explicitly estimating CSI for each node. The algorithm proceeds as follows. In each iteration, transmitters perturb their phase randomly and independently from a predefined distribution and transmit a common signal. The receiver measures the combined response and broadcasts a single bit of feedback. The feedback bit $F[t]$ that is sent at the end of time slot $t$ indicates the result of comparison of the current measured RSS with the highest RSS observed over a moving window of the last  $M$ iterations:
\begin{equation}
F[t]=\left\{ \begin{array}{ll}
1 & \textnormal{if } {R[t]>R_\text{best}[t]} \\
0 & \textnormal{if } {R[t]<R_\text{best}[t]} \\
\end{array} \right.
\label{eq:1bf_RX},
\end{equation}
where 
\begin{equation}
R_\text{best}[t]=\smash{\displaystyle\max_{t-M \leq \tau<t-1}} \; R[\tau]
\label{eq:1bf_window}.
\end{equation}
Each transmitter updates its phase according to the feedback from the receiver by keeping the random perturbation for a positive feedback and discarding it in the case of negative feedback:

\begin{equation}
\theta_{i}[t+1]=\left\{ \begin{array}{ll}
\theta_{i}[t]  & \textnormal{if } F[t]=0 \\
\theta_{i}[t]+\delta_i[t] & \textnormal{if } F[t]=1 \\
\end{array} 
\label{eq:1bf_TX} \right..
\end{equation}
where $\delta_i[t]$ is the phase perturbation added in iteration $t$.

\subsubsection{Randomized two bit feedback algorithm (R2BF)}

The authors in \cite{thibault2010random} propose a modified version of OBF, namely the randomized 2 bit feedback algorithm (R2BF), in order to speed up convergence. Assuming that the receiver has knowledge of the maximum possible RSS value obtained by perfect beamforming, the feedback bits are set as follows:
\begin{equation}
F[t]=\left\{ \begin{array}{ll}
11 & {\textnormal{if } R[t]>R_\text{best}[t] \textnormal{ and close to  RSS}_\text{max}} \\
10 & {\textnormal{if } R[t]>R_\text{best}[t] \textnormal{ half way from  RSS}_\text{max}}\\
01 & {\textnormal{if } R[t]>R_\text{best}[t] \textnormal{ and far from  RSS}_\text{max}} \\
00 & {\textnormal{if } R[t]<R_\text{best}[t]} \\
\end{array} \right.
\label{eq:R2BFeq}
\end{equation}
and the phase update (\ref{eq:1bf_TX}) becomes
$$
\theta_{i}[t+1]=\left\{ \begin{array}{ll}
\theta_{i}[t] &\textnormal{if } F[t]=00 \\
\theta_{i}[t]+\delta_i[t] & \textnormal{otherwise} \\
\end{array} \right.
$$
In the next time slot, $\delta_i[t+1]$ is chosen from a different distribution  depending on the feedback, i.e. $\delta_i[t+1] \sim U[-\frac{\pi}{\beta},\frac{\pi}{\beta}]$ where:
$$
\beta=\left\{ \begin{array}{ll}
\beta_1 & \textnormal{if } F[t] = 01 \\
\beta_2 & \textnormal{if } F[t] = 10 \\
\beta_3 & \textnormal{if } F[t] = 11 \\
\end{array} \right.
$$
where $\beta_1<\beta_2<\beta_3$. This approach increases the convergence speed of 1BF by around 25\%, but with the additional requirement of the receiver knowing the maximum RSS, which in turn requires knowledge of the number of transmitters and the channel statistics. This method therefore requires a higher level of coordination between nodes and is less robust and distributed.

\subsubsection{Modified two bit feedback algorithm (M2BF)}

We propose a different modification of one bit feedback, where the additional bit of feedback is used to quantify the {\it amount} of improvement obtained from the perturbations. 
The additional feedback bit relative to Eq. \ref{eq:1bf_RX} is generated as follows:
\begin{equation}
F[t]=\left\{ \begin{array}{ll}
11 & {\textnormal{if } \alpha_1 \leq R[t]-R_\text{best}[t] } \\
10 & {\textnormal{if } 0 \leq R[t]-R_\text{best}[t] <\alpha_1}\\
01 & {\textnormal{if } \alpha_2 \leq R[t]-R_\text{best}[t] <0}\\
00 & {\textnormal{if } R[t]-R_\text{best}[t] \leq \alpha_2 }\\
\end{array} \right.
\label{eq:2bf_TX}
\end{equation}
where $\alpha_1$ and $\alpha_2$ are predefined constants dependent on channel statistics, but independent of $N$. If the RSS {\it improvement} from current random phase perturbations is above threshold $\alpha_1$, all transmitters make use of this knowledge and apply the previous perturbations again in the next iteration. If the {\it degradation} caused by the perturbation is more than threshold $\alpha_2$, the phases are reversed and transmitters perturb their phases in the opposite direction. Therefore $\delta_i[t+1]$
becomes dependent on the previous perturbation $\delta_i[t]$ as follows:
\begin{equation}
\delta_i[t+1]=\left\{ \begin{array}{ll}
 \textnormal{new random}& {\textnormal{if } F[t] = 10 \textnormal{ or } 01 } \\
\delta_i[t]& {\textnormal{if } F[t] = 11}\\
-\delta_i[t] & {\textnormal{if } F[t] = 00}\\
\end{array} \right.
\label{eq:2bf_TX_decision}
\end{equation}

\subsection{Numerical results and comparisons}

To evaluate and compare the beamforming performance 
and convergence speed of the preceding algorithms, we fix the number of feedback bits to 2 and plot the progression of each algorithm with the number of iterations.
We compare DOST with 2-bit quantized feedback against the 2-bit SDDB, M2BF and R2BF strategies discussed above. 
While our later system-level numerical results are for a DBS with 10 nodes, in this section, we consider a larger number of nodes ($N=100$) in order to stress test
the feedback strategies considered. We investigate the evolution of beamforming gain as a function of iterations using Monte Carlo simulations.

The distributed beamforming schemes used in the simulations have the following parameters: the phase perturbations $\delta_i[t]$ are generated from the uniform distribution $\mathcal{U}(-10^\circ,10^\circ)$. R2BF uses $\beta_1 = 5^\circ$, $\beta_2 = 10^\circ$, $\beta_1 = 25^\circ$ and constant thresholds of $\xi_1 =0.3$, $\xi_2=0.8$ to decide from the $R_\text{best}$ where RSS fits in (\ref{eq:R2BFeq}). For M2BF, $\alpha_{1,2} = 0.8$ and the phase perturbations are designed to decay exponentially from $\mathcal{U}(-45^\circ,45^\circ)$ with the number of iterations.

\begin{figure}[h]
\centering
\includegraphics[width=0.9\columnwidth]{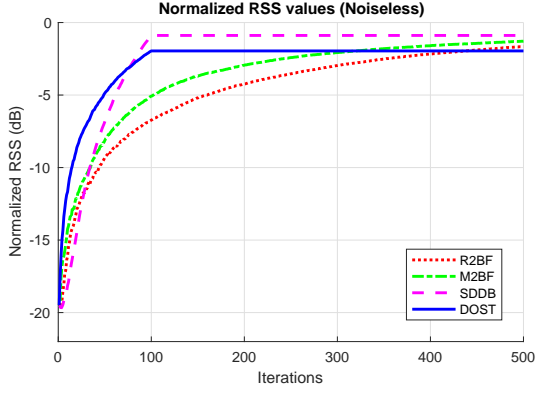}
\caption{Evolution of the distributed beamforming approaches in a noiseless setting for $N = 100$ nodes}
\label{fig:evolution_all_noiseless}
\end{figure}

We first focus on understanding the effect of quantization in a noiseless setting.
Figure \ref{fig:evolution_all_noiseless}  shows, at each iteration, the RSS level that would be obtained by nodes using their current channel estimate for beamforming. The curves of Figure \ref{fig:evolution_all_noiseless}  are the result of averaging over 2000 realizations of an $N=100$ element array in a noiseless setting. For deterministic algorithms, training is stopped after $L=100$ iterations, which constitutes one ``batch" of training. In the absence of noise, the stochastic algorithms, R2BF and M2BF, converge asymptotically to optimal beamforming, while deterministic algorithms hit an performance gap of 1 and 2 dB away from optimal beamforming for SDDB and DOST, respectively, due to feedback quantization.
When the feedback link is not a bottleneck, increasing the number of feedback bits can also be used to decrease quantization loss, but our interest is in the low per-node SNR
regime, where this is not a feasible strategy.  Thus, in a noiseless setting with severe feedback quantization, the one-bit feedback algorithm and its variants actually perform better
than deterministic training.  And among the deterministic strategies, time-multiplexing across nodes as in SDDB is better than DOST, since the dynamic range of
the received signal is smaller.

\begin{figure}[h]
\centering
\includegraphics[width=0.9\columnwidth]{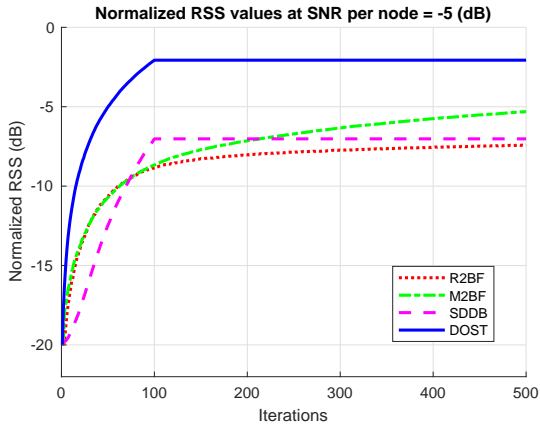}
\caption{Evolution of the distributed beamforming approaches for $N = 100$ nodes at SNR per node $-5$ dB}
\label{fig:evolution_all}
\end{figure}

However, the picture is quite different when we consider the low per-node SNR regime of interest to us. Figure \ref{fig:evolution_all} shows the evolution of the beamforming algorithms at per-node SNR of $-5$ dB. Since the SDDB scheme does not get the benefit of time averaging, it falls 7 dB short of the ideal 20 dB beamforming gain after $N$ iterations, whereas DOST comes to within 2 dB of the ideal beamforming gain.
The one-bit beamforming schemes perform very poorly and we do not plot it.  Among its two-bit variants, the M2BF scheme performs better than the R2BF scheme, but falls well short of
the ideal beamforming gain: 8 dB lower after $N=100$ iterations, and 5 dB lower even after 500 iterations.  Thus, the DOST algorithm is by far the most resilient at low per-node SNR.

\subsubsection*{Impact of quantization}

\begin{figure} [t]
\centering
\includegraphics[width=0.9\columnwidth]{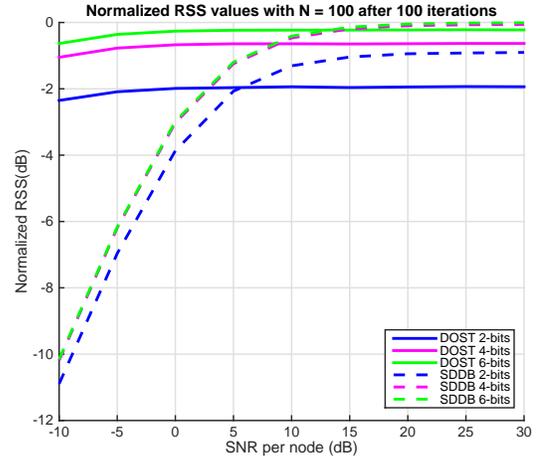}
\caption{Normalized beamforming gains for SDDB and DOST for different number of feedback bits}
\label{fig:SDDBvsDOST_feedback_bits}
\end{figure}

We now explore the impact of quantization further, by varying the number of bits of feedback per iteration in the DOST and SDDB algorithms. The number of feedback bits is fixed to 2 for M2BF and R2BF, hence
we do not consider those schemes here.
Figure \ref{fig:SDDBvsDOST_feedback_bits} shows the performance of the two algorithms with different levels of feedback quantization for different per-node SNR after 100 iterations of training.
We consider 2-bit feedback quantization to quantize both real and imaginary parts of the received baseband signal. With 2-bit quantization, DOST is able to achieve within 2 dB  of the ideal solution. Increasing the number of quantization bits to 4 bits improves both algorithms by around 1 dB and further increasing it to 6 bits gives very slight performance improvement. These results show that DOST  can achieve near-optimal beamforming gains with heavily quantized feedback, as low as 2 bits per iteration, making it competitive with stochastic ascent approaches like R2BF and M2BF, even in noise-free conditions where they perform best. 

\section{OFDM Pilot Design \label{sec:wideband}}

\begin{figure} [t]
\centering
\includegraphics[width=0.6\columnwidth]{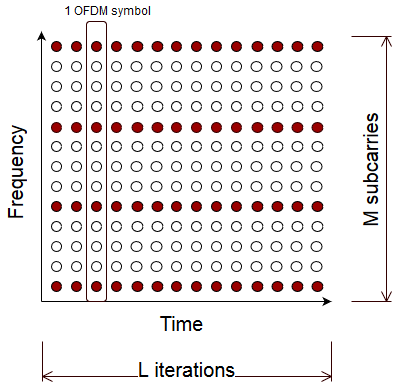}
\caption{Proposed comb type pilot arrangement for channel estimation over OFDM grid}
\label{fig:block_type}
\end{figure}

To extend the framework to wideband, an OFDM framework is utilized wherein DOST is applied on a subset of the OFDM subcarriers by placing training pilots at
 known positions in the OFDM symbol grid. Different pilot placements are possible for the training, including the block type, the comb type, or 2D-grid type \cite{colieri2002study}. In a block type arrangement, the pilots are placed on all subcarriers in a few OFDM symbols; in the comb type, the pilots are present in all OFDM symbols over a subset of subcarriers as shown in Figure \ref{fig:block_type}; and in the 2D-grid type, the pilots are present in a subset of OFDM symbols over a subset of subcarriers. Therefore, the number of pilots in the 2D-grid pattern are less than the block and comb type pilot arrangements. 
Our goal here is to learn the channel as quickly as possible, hence for any given subcarrier, it is best to concentrate our pilot resources in time (over $L \geq N$ successive OFDM symbols for an $N$-node DBS) so as to get the required feedback from the receiver as quickly as possible. This is particularly important for maximizing the rate of channel time variations a DBS can support, because of the relatively low rate of feedback available on the uplink (see Section \ref{sec:outage}). However, by exploiting the continuity of the channel across frequency, we only need to employ pilots for a subset of subcarriers, and estimate the optimum beamforming weight for all other subcarriers via interpolation in the frequency domain.  We therefore consider the comb type pilot arrangement shown in Figure \ref{fig:block_type} for the DBS deployment.

A number of different interpolation strategies can be used to extend the pilot subcarrier channel estimates to the remaining subcarriers, including linear interpolation, second order interpolation, low pass interpolation, spline cubic interpolation, and  time domain interpolation.  We consider lowpass interpolation, which has been shown
to work better with comb type pilots \cite{colieri2002study}.

\begin{figure} [h]
\centering
\includegraphics[width=0.6\columnwidth]{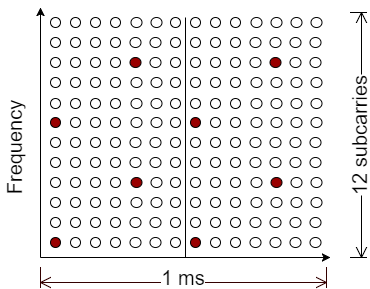}
\caption{Standard 2D-grid type pilot placement is not well matched to a DBS OFDM downlink.}
\label{fig:LTE_2D}
\end{figure}

The typical LTE system parameters shown in Table \ref{table:LTE} are used in the simulations; and comb type pilots are placed at all OFDM symbols on a subset of 200 equispaced subcarriers. The minimum required number of OFDM symbols for training is $L \geq N$, and the minimum required time for training is 0.71 ms for $N=10$ nodes.
Note that a standard 2D grid type pilot arrangement, as illustrated in Figure \ref{fig:LTE_2D}, would require a longer training time.
For the system parameters of Table \ref{table:LTE}, the minimum required number of OFDM symbols for training is also $L \geq N$, however, a subset of OFDM symbols are used as the pilots and the required time for training is $7L$ OFDM symbols which corresponds to a minimum training time of 5 ms for $N=10$ nodes. This gap grows linearly with the number of nodes and can become a bottleneck when scaling to larger arrays. 

\begin{table}[h!]
\begin{center}
\begin{tabular}{ |c| |c|  }
 \hline
 Variables& Parameters\\
 \hline
 Number of nodes (N) & 10 \\ \hline
 Bandwidth (Downlink)   & 20 MHz  \\ \hline
 Bandwidth (Uplink)   & 20 MHz  \\ \hline
 Number of subcarriers&   1200\\ \hline
 Number of pilot subcarriers&   200\\ \hline
 Size of FFT & 2048\\ \hline
 Subframe length&1 ms\\ \hline
 OFDM symbols per subframe&14\\ \hline
 Channel Model& EPA\\ \hline
 Doppler Spread& 5 Hz\\ 
 \hline
\end{tabular}
\end{center}
\caption{Link Level System parameters}
\label{table:LTE}
\end{table}

\subsection{Numerical results \label{sec:feedbacklink}}

\begin{figure} [t]
\centering
\includegraphics[width=0.9\columnwidth]{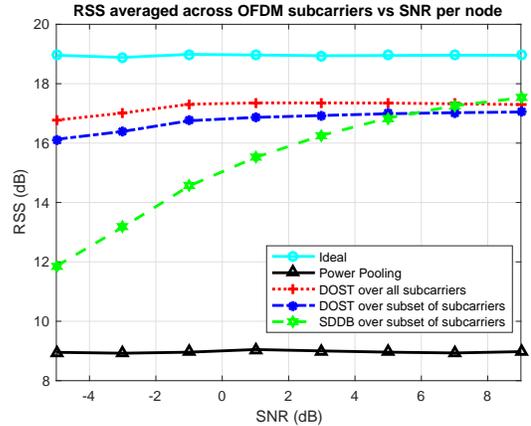}
\caption{RSS averaged over OFDM subcarriers vs SNR per node for $N=10$ nodes}
\label{fig:wideband}
\end{figure}

Simulation results using uncorrelated EPA channels with the power delay profile of Table-\ref{table:EPA}
and system parameters of Table-\ref{table:LTE} are reported here for the wideband setting. The comb type pilot arrangement is used with subcarrier spacing of $6$.
Figure \ref{fig:wideband} shows the performance of beamforming algorithms averaged over subcarriers for $N=10$ nodes. DOST achieves within 3 dB of the ideal solution, and works better than SDDB at low per-node SNR. The degradation due to interpolation of comb type pilots is only 0.5 dB compared to a setting in which all subcarriers are used as pilots.

\section{System Performance \label{sec:outage}}

We now provide performance characterization via outage rates for the downlink transmission of pilots and data and for feedback broadcast in the uplink direction.  For simplicity, 
we use the terms ``capacity'' and ``outage rate/capacity'' to denote spectral efficiencies, either for a narrowband system (modeling a single subcarrier), or averaged over subcarriers for a wideband system.  We use the term ``data rate'' when we multiply such spectral efficiencies by the bandwidth.  

The forward link enjoys the benefits of $N$-fold power pooling gain during training, and $N^2$-fold distributed transmit beamforming gain during post-training data transmission.  The rate of the feedback link depends on the sophistication of its reception strategy, as well as its allocated resources, which may be less than that of the forward link. There is no power pooling on the uplink and transmission is at an $N$-fold disadvantage in this direction relative to downlink. 
This asymmetry may be offset by using distributed receive beamforming in the feedback direction to pool the resources of the array and allow uplink scaling to keep up with downlink. Different analog and digital receive beamforming algorithms have been proposed in previous works such as \cite{quitin2016scalable,brown2014distributed}.
In the worst case, however, feedback is delivered over a SISO channel, either by having each node decode and use the feedback independently or use a single designated node for feedback reception.
In this case, if the downlink power emitted by a single node in the DBS cluster and the uplink power emitted by the user node are comparable, then the feedback link may well be the scaling bottleneck and necessitate longer symbol durations to build up SNR, which will limit the rate of channel time variations that can be supported by the distributed array. 

\begin{figure*} 
	\includegraphics[width=\textwidth]{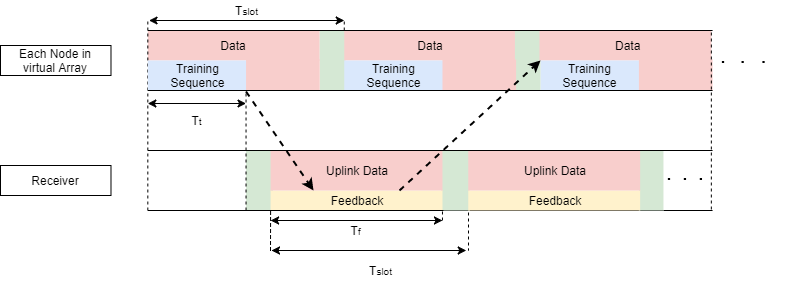}
	\caption{DBS frame structure with a slow feedback link.}
\label{fig:frame_structure}
\end{figure*}

Figure \ref{fig:frame_structure} illustrates an example frame structure.
During the startup phase, data could be sent in power pooling mode at lower spectral efficiency, while sending training on the pilot subcarriers.  Beamforming can be applied once the feedback corresponding to the pilots has been received.  Once continuous communication has been established, 
feedback regarding the designated set of pilot subcarriers on the downlink in frame $f$ is sent back during the next frame $f+1$, and the results
are applied for distributed transmit beamforming in frame $f+2$.    With such a scheme, for the first and second frames, data could still be sent in power pooling mode at lower spectral
efficiency, with distributed beamforming enabled from the third frame onwards.  For continuous communication, we would be beamforming in a given frame based on pilots sent two frames back. If the feedback link is the bottleneck, then the frame length $T_f$ must be long enough to carry back the feedback corresponding
to $M_p$ pilot subcarriers on $L \geq N$ OFDM symbols, which amounts to $2 M_p L$ bits using 2-bit feedback.  For feedback of rate $R_f$, this corresponds
to a frame length of $T_f = \frac{2 M_p L}{R_f}$, and a channel coherence time of $T_c = 3 T_f$.  If the feedback link is not the bottleneck, then the minimum frame length can be set to $T_f = L T_\text{OFDM}$, where $T_\text{OFDM}$ is the OFDM symbol length, which is governed by the channel delay spread and the overhead allowed
for the cyclic prefix.  We provide numerical values for our running example of a 20 MHz downlink over an EPA channel model at the end of this section.

We first discuss achievable performance on the downlink, and then consider the feedback link. In both cases, we use outage rates for a compact bottom-line
characterization.

\subsection{Downlink Performance} \label{sec:downlink}

We assume that each node in the DBS cluster applies a phase correction on each subcarrier based on its channel estimate, and employs a uniform power distribution across subcarriers. Since the DBS nodes do not have information regarding the relative channel strengths across subcarriers or transmitters, more optimal power allocation methods such as waterfilling are not possible.
Even for a narrowband channel, using multiple transmitters for distributed beamforming provides spatial diversity, hence we derive a pessimistic estimate of outage spectral efficiency
by ignoring frequency diversity and focusing on  
a single subcarrier at frequency $f_k$. The channel seen by node $i$ is denoted as $H_i (f_k)$. Upon ideal phase compensation, the net channel seen at the
receiver is given by $\|{\bH(f_k)}\|_1 = \sum_{i=1}^{N} |{H_i(f_k)}|$, where $\bH(f_k) = (H_1 (f_k) ,...,H_N(f_k))^T$ is the vector of channel gains corresponding to the $N$ nodes in the DBS cluster. Modeling the channels $\{ H_i (f_k ) \}$ as zero mean complex Gaussian normalized as $E[ |H_(f_k)|^2 ] = 1$, the effective channel amplitude gain $\|{\bH(f_k)}\|_1$ is a sum of i.i.d. Rayleigh random variables, each with mean squared value of one.

Assuming that each transmitter applies power $P$ to each subcarrier, the outage probability for a narrowband system operating at $f_k$ is given by
\begin{equation}
\begin{aligned}
p_\text{out}(R) &= \mathbb{P} \bigg\{ \log_2 \bigg(1 + \frac{P \|{\bH(f_k)}\|_{1}^2}{N_0} \bigg) < R \bigg\} 
\\
&= \mathbb{P}\bigg\{ \|{\bH(f_k)}\|_1 < \sqrt{(2^R-1)\tfrac{N_0}{P}} \bigg\}
\label{outage_rate}
\end{aligned}
\end{equation}
The $\epsilon$-outage capacity $C_{\epsilon}$ is the maximum rate $R$ such that $p_\text{out}(R)$ is less than $\epsilon$.

Letting $F( \cdot )$ denote the CDF of $\|{\bH(f_k)}\|_1$, we see that
\begin{equation} \label{eq:Ceps}
C_{\epsilon}=  \log_2 \bigg( 1+  \frac{P}{N_0} F^{-1}(\epsilon)^2 \bigg)
\end{equation}
Since $\|{\bH(f_k)}\|_1 = \sum_{i=1}^{N} |{H_i(f_k)}|$ is a sum of i.i.d. random variables, we get insight, and a good approximation, by
applying the central limit theorem. That is, we can approximate $\|{\bH(f_k)}\|_1$ as Gaussian with mean $\mu = N \sqrt{\nicefrac{\pi}{4}}$ and variance $\sigma^2 = N (1-\nicefrac{\pi}{4})$.
Using this approximation in (\ref{eq:Ceps}), we obtain that 

\begin{equation}
C_{\epsilon} \approx \log_2\bigg( 1+ \frac{P}{N_0} \bigg(N \sqrt{\tfrac{\pi}{4}} - \sqrt{N \tfrac{(1-\pi)}{4}} Q^{-1}(\epsilon)  \bigg)^2 \bigg)
\label{eq:Gaussian}
\end{equation}
where $Q( \cdot )$ denotes the complementary CDF of a standard Gaussian random variable.  This indicates that that the outage capacity shows a $\log N$ growth with the number
of nodes, with $O( \sqrt{N} )$ backoff within the argument of the logarithm in order to handle the tails. 

The Gaussian approximation works well for moderately large $N$, including our running example of $N=10$, and provides insight into the benefits of both spatial diversity and beamforming.
We note, however, that for small $N$, the outage capacity approximation can be improved by using a small argument approximation to the CDF $F$ of
a sum of i.i.d. Rayleigh random variables \cite{hu2005accurate}, given by
\begin{equation} \label{eq:SAA}
\begin{aligned}
F_\text{SAA}(t\sqrt{N}) &\approx 1 - e^{-\frac{t^2}{2b}} \sum_{k=0}^{N-1} \frac{(\frac{t^2}{2b})^k}{k!}
\\ 
b &= \frac{\sigma^2}{N} \big[ \prod_{i=1}^{N}(2i-1) \big]^{1/N}
\end{aligned}
\end{equation}
where $t = \frac{x}{\sqrt{N}}$ is a normalized argument for the CDF.
This approximation, when used in (\ref{eq:Ceps}),  is excellent for small values of $t$ which is the regime of interest for the outage probabilty $\epsilon$.

We compare these approximations with simulations in the next section.

\subsubsection*{Numerical results}

\begin{figure} [t]
\centering
\includegraphics[width=0.9\columnwidth]{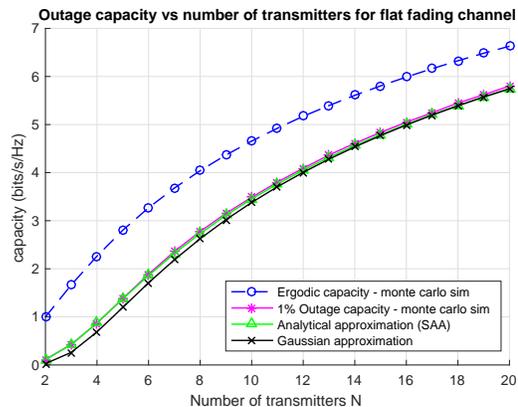}
\caption{Ergodic capacity and 1\% outage rates (b/s/Hz) versus number of nodes for narrowband flat fading channel at SNR $= -5$ dB}
\label{fig:outage_capacity_narrowband_-5dB}
\end{figure}

Figure \ref{fig:outage_capacity_narrowband_-5dB} shows the ergodic capacity and the outage rate versus the number of transmitters at $-5$ dB SNR per node for a narrowband channel with ideal channel state information. The ergodic capacity and the $1\%$ outage rate curves are obtained with Monte Carlo simulations.
The analytical outage capacity approximation for sum of Rayleigh random variables in (\ref{eq:SAA}) matches Monte Carlo simulations very well and the Gaussian aprroximation of the sum of Rayleigh random variables (\ref{eq:Gaussian}) is slightly pessimistic for the small number of nodes.   The difference between ergodic capacity and outage rate diminishes as the number of nodes increases because the diversity gain provided by multiple nodes reduces the variance of the aggregate channel and, in turn, the variance of spectral efficiency. It can be observed that, with $N=10$ nodes, the outage capacity of $3.5$ bps/Hz can be obtained at $-5$ dB SNR per node.

\begin{figure} [t]
\centering
\includegraphics[width=0.9\columnwidth]{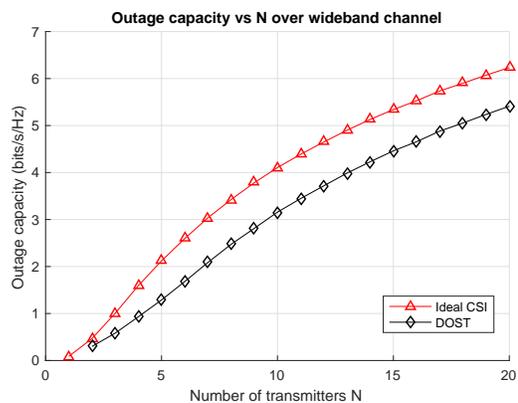}
\caption{1\% outage rates (b/s/Hz) vs N for wideband channel at average SNR $= -5$ dB}
\label{fig:outage_capacity_wideband}
\end{figure}

Figure \ref{fig:outage_capacity_wideband} shows Monte Carlo simulation results for outage capacity versus number of transmitters applied to the wideband setting (i.e., where the spectral efficiency is averaged over the signal bandwidth) with parameters in Table-\ref{table:LTE} at $-5$ dB average SNR. The ideal CSI curve shows the capacity when the channel is known to all nodes and perfect beamforming is applied over the entire frequency band. The DOST curve shows Monte Carlo simulation results with 2 bits of feedback per pilot subcarrier.  The heavily quantized DOST algorithm provides significant gains in terms of capacity and is able to achieve outage rate of $3.1$ bps/Hz using 10 nodes. 
Thus, even while operating at a per-node SNR of $-5$ dB, DOST can yield a data rate of about 50 Mbps over a 20 MHz band, after accounting for the overhead of reserving $1/6$ of the subcarriers for pilots, under the assumption that we would like to be as reactive to channel time variations as possible and therefore insert comb type pilots into every OFDM symbol. For transmit power of 20 dBm (100 mW) per DBS node, isotropic antennas, and receiver noise figure of 6 dB, the attainable range using the Hata propagation model at 800 MHz carrier frequency with 30 m DBS node height and 1.5 m receiver height is about 6.6 km, allowing for a 5 dB implementation margin (we already account for fading in our formulation, hence we do not require excess link margin to accommodate it). The range that can be attained in the same setting for the same target data rate is 2.3 km for a single node and 4.2 km with power pooling.
As expected, the corresponding numbers at a lower carrier frequency of 200 MHz are better: 5 km for a single node, 9 km with power pooling, and 14.3 km with DOST.

\subsection{Feedback link}

While we do not consider detailed design of the feedback link, we provide insight into the impact of reception strategy by comparing three different options.
The first and simplest approach is to designate a single node
(e.g., one of the DBS nodes) as receiver for the feedback.  In this case, the received SNR is very low (e.g., $-5$ dB for our running
example, assuming that the emitted power from a DBS and user node are similar), and the feedback link becomes a significant bottleneck.
A second approach is to attempt to decode the feedback packet at each DBS node
separately, and to assume that successful decoding at any node will enable all other nodes to obtain the feedback via broadcast on a fast local area network (LAN).
The third, and most complex, is distributed receive beamforming. Digitization and local transmission of the received signals at the $N$ DBS nodes to a centralized processor
requires that the LAN speed scale with $N$. It has been shown in \cite{brown2014distributed} that much of the received beamforming gains (within 2 dB of ideal) can be obtained even if hard decisions
are exchanged: this still requires LAN speed scaling with $N$, but at a smaller rate. Amplify-forward approaches for receive beamforming which sidestep such local communication
by enabling on-air combining have also been proposed and demonstrated \cite{quitin2016scalable}.  We therefore consider ideal receive beamforming as providing a performance benchmark for
the feedback channel that may be attainable with sufficient engineering effort.  

The bandwidth on the feedback link may be different (typically smaller, since multiple user nodes may be sending feedback to the DBS) from that on the downlink.  We average
the spectral efficiency across this bandwidth when determining outage rates.

In the first approach, for a discrete set of $M_u$ subcarriers, the spectral efficiency at a given DBS node, say node $k$, can be calculated as
\begin{equation}
I_k = \frac{1}{M_u} \sum_{i=1}^{M_u} \log_2(1+ \textnormal{SNR} |H_k(f_i)|^2)
\label{eq:spectralEfficiecy_approx}
\end{equation} 
where $H_k (f_i)$ is the uplink channel on the $i$th subcarrier for the $k$th DBS node.
We can now define the $\epsilon$-outage rate $R_u$ as usual
\begin{equation}
P(I_k <R_1)= \epsilon .
\label{eq:outage_cap1}
\end{equation} 

For the second approach,  outage occurs if all of the DBS nodes are unable to decode the feedback packet:
\begin{equation}
P\left( max(I_1,I_2,...,I_N) <R_2 \right) = \epsilon .
\label{eq:outage_cap2}
\end{equation} 
Assuming that the channel realizations for the different nodes are i.i.d., we infer that the random variables $I_1,...,I_N$ are i.i.d., so that
$P\left( max(I_1,I_2,...,I_N) <R_2 \right) = \left( P(I_1 < R_2) \right)^N$.  Thus, we obtain that the outage rate satisfies
\begin{equation}
P\left( I_1 <R_2 \right) = \epsilon^{\frac{1}{N}} .
\label{eq:outage_cap2b}
\end{equation} 
which allows the individual outage probability at any DBS node to be much higher.  We have checked via simulations that
(\ref{eq:outage_cap2}) and (\ref{eq:outage_cap2b}) yield the same results for our channels, which are obtained by independent draws from
the EPA  model.

For the third approach (receive beamforming), the spectral efficiency is given by
\begin{equation}
I_\text{beam} = \frac{1}{M_u} \sum_{i=1}^{M_u} \log_2 \left(1+ \textnormal{SNR} (\sum_{k=1}^N |H_k(f_i)|^2) \right)
\end{equation} \label{eq:rx_beam}
and the outage rate satisfies
\begin{equation}
P\left( I_\text{beam} <R_3 \right) = \epsilon .
\label{eq:outage_cap3}
\end{equation}

\begin{figure} [t]
\centering
\includegraphics[width=0.9\columnwidth]{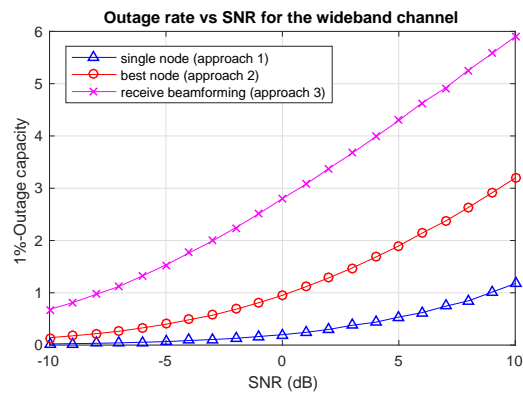}
\caption{1\% outage rates vs per node SNR for wideband channels using 10\% of the 20 MHz uplink bandwidth with single node, best node and ideal receive beamforming with using $N=10$ nodes}
\label{fig:uplink_outage_capacity_wideband}
\end{figure}

Figure \ref{fig:uplink_outage_capacity_wideband} shows the outage rates for the three approaches computed over a $W_\text{fb} = 2$ MHz bandwidth for the feedback link.  For a per-node SNR of $-5$dB, the outage rates are
given by $R_1 = 0.066$, $R_2 = 0.4$, and $R_3 = 1.5$ bps/Hz.  These translate to feedback data rates $R_f$ of 132 Kbps, 0.8 Mbps, and 3 Mbps, respectively. Assuming that the feedback link is the bottleneck, and that $L=N$, we compute the corresponding frame lengths as
\begin{equation}
T_f =  \frac{2 N M_p}{R_f}
\label{eq:Tf}
\end{equation}
The minimum channel coherence times $T_c = 3 T_f$ that can be supported using the three approaches are given by
90, 15 and 3.9 ms respectively.  

We conclude that, even though the SNR on the feedback link is so low, the first approach is adequate for quasi-static links typical of rural broadband.
However, if more sophisticated strategies such as the second or third approaches are employed, the DBS concept can be used to support moderate mobility.

\section{Conclusion \label{sec:conclusion}}

The distributed base station concept presented in this paper is a promising approach for providing broadband access to remote areas, combining the benefits of massive MIMO
with the superior propagation characteristics of large carrier wavelengths (e.g., white space frequencies).  While distributed transmit beamforming
is an intuitively plausible approach for range extension, we have identified and addressed key technical challenges in realizing this potential.  First, we observe that 
the DBS system is inherently a low-SNR configuration, so that noise-resilient algorithms are required to realize the gains of distributed beamforming for this application. 
We demonstrate that algorithms that utilize deterministic or pseudo-random beamforming sequences over a designated training phase greatly outperform stochastic ascent algorithms such as the 1-bit feedback scheme and its variants. Among the deterministic algorithms, the proposed distributed orthogonal sequence training (DOST) scheme outperforms per-node training due to its capability of aggregating the training measurements of the entire array and boosting SNR through noise averaging.  The DOST algorithm, by bootstrapping from power-pooling gain, scales indefinitely with the size of the distributed array, at the expense of linear growth in training overhead.  Second, we have shown that narrowband beamforming strategies extend naturally to the wideband regime by training on a subset of pilot subcarriers, and then interpolating across subcarriers. 
Third, we observe that the feedback link can become a bottleneck if distributed receive beamforming is not employed, and
show that this impacts the rate of channel time variations that the distributed array can track.  However, we show that it is possible to support relatively slowly varying links (e.g., with coherence times of the order of 100 ms) even in this setting, and that faster channel time variations can be supported with more sophisticated uplink reception strategies.

The promising results in this paper motivate efforts to prototype and experimentally demonstrate range extension with the DBS concept.
The identification of the feedback link as the bottleneck motivates serious investigation of more sophisticated distributed reception strategies 
in the low per-node SNR regime.

\section*{Acknowledgments}
This research was supported by the National Science Foundation under grant CCF-1302114.
\bibliographystyle{IEEEtran}

\bibliography{references}

\end{document}